# Semantic Query Integration With Reason

Philipp Seifer[a], Martin Leinberger[b], Ralf Lämmel[a], and Steffen Staab[b,c]

a Software Languages Team, University of Koblenz-Landau, Germany

b Institute for Web Science and Technologies, University of Koblenz-Landau, Germany

c Web and Internet Science Research Group, University of Southampton, England

**Abstract** Graph-based data models allow for flexible data representation. In particular, semantic data based on RDF and OWL fuels use cases ranging from general knowledge graphs to domain specific knowledge in various technological or scientific domains. The flexibility of such approaches, however, makes programming with semantic data tedious and error-prone. In particular the logics-based data descriptions employed by OWL are problematic for existing error-detecting techniques, such as type systems. In this paper, we present DOTSpa, an advanced integration of semantic data into programming. We embed description logics, the logical foundations of OWL, into the type checking process of a statically typed programming language and provide typed data access through an embedding of the query language SPARQL. In addition, we demonstrate a concrete implementation of the approach, by extending the Scala programming language. We qualitatively compare programs using our approach to equivalent programs using a state-of-the-art library, in terms of how both frameworks aid users in the handling of typical failure scenarios.

## 1 Introduction

Graph-based data models allow for flexible data representation. In particular, semantic data models like RDF [55] may contain schematic information as part of the data or in separate files. Such schematic information is called an ontology. Of special interest is the W3C standard OWL [25], which allows for using highly expressive logic-based data descriptions. This flexibility and expressive power of RDF and OWL fuels many applications, ranging from general knowledge graphs such as Wikidata [54] to complex domain specific ontologies, e.g., the SNOMED CT [8] medical vocabulary.

While the flexibility and expressive power of OWL make it attractive, programming with OWL is tedious and error-prone. A major reason is the lack of typed integration in programming languages, leaving the burden of correct typing on the programmer. This lack of integration is comparable to data access and integration of other data models, such as access to and types for relational [6] or object oriented databases [37, 56], XML [4, 26], as well as general data access approaches such as LINQ [5, 34]; each data model comes with its specific challenges. As an example of these challenges, consider the following axioms inspired by the Lehigh University Benchmark [23]:

```
1  // Schematic information
2  Person ⊓ Organization ⊑ ⊥
3  Employee ⊑
4      Person ⊓ ∃worksFor.Organization
5  Professor ⊑ Employee
6  Chair ⊑ Professor
7  ∃headOf.Department ⊓ Person ≡ Chair
8  ResearchAssistant ⊑
9      Person ⊓ ∃worksFor.ResearchGroup
10 Department ⊑ Organization
11 ResearchGroup ⊑ Organization
12 ∃worksFor.⊤ ⊑ Person
13 ∃subOrganizationOf.⊤ ⊑ Organization
14 ⊤ ⊑ ∀headOf.Department
15 // Data (assertional axioms)
16 alice : Chair
17 (bob, softlang) : worksFor
18 softlang : ResearchGroup
```

**Figure 1** Example axioms describing a university setting.

Schematic information consists of concepts, such as *Person*, and roles, such as *worksFor*, that are combined to more complex concept expressions via connectors like intersection (⊓) or existential (∃) and universal (∀) quantification. Concepts themselves are related to each other via subsumption (⊑) or equivalence (≡) statements. In this manner, line 2 ensures that a *Person* can never be an *Organization* and vice versa. Lines 3–4 state that *Employees* are *Persons* working for some kind of *Organization*. Line 5 introduces *Professors*, which are *Employees*. A special kind of a *Professor* is a *Chair* (6) which is a person who is the head of a department (7). A *ResearchAssistant* is a person that works for a *ResearchGroup* (8–9). Both *Departments* and *ResearchGroups* are special kinds of *Organizations* (10,11). To express that working for something implies being a *Person* and being a sub-organization implies being an *Organization*, domain specifications are employed (12,13). The range specification (14) ensures, that all objects to which a *headOf* relation points are *Departments*. The data introduces three objects: *alice*, who is a *Chair*, and *bob* who works for the *ResearchGroup* *softlang*.

This example highlights some of the problems that occur when trying to do type checking on a program that works on OWL. For one, a mixture of nominal (*Person*) and structural types ($\exists$*worksFor.Organization*) is used. Second, some inferable information is left implicit, such as the fact that a *ResearchAssistant* is a special kind of *Employee*.

Mapping approaches such as [27] do not cope well with these problems. In previous work [30], we proposed a custom type system dubbed $\lambda_{DL}$ to remedy the situation. $\lambda_{DL}$ used concept expressions such as $\exists$*worksFor.ResearchGroup* or *Employee* as types. The process of type checking then relies on an ontology reasoner. This allows for the definition of functions, e.g., a function accepting an *Employee* such as ($\lambda x$ : *Employee* . ...), as well as proving that a *ResearchAssistant* is a subtype of *Employee* through the reasoner. Besides the possibility of finding wrong applications of this function at compile time, this also serves as documentation, which is guaranteed to be consistent [42] with both the program code as well as the ontology.

A practical integration of OWL into programming, however, must extend general purpose programming languages. Those have rich type systems, where even small changes may be cross-cutting among many features. In addition, expressive queries as a means to access data are needed. In particular, a typed integration of SPARQL [43], a W3C querying standard for semantic data, is desirable.

In this paper, we describe a general approach for a deep integration of OWL and a subset of SPARQL into a typed programming language as well as a concrete implementation, ScaSpa, as an extension of Scala. The subset of SPARQL we consider is built on [28] in order to focus on SPARQL constructs that are decidable when used with OWL. In summary, the **main contributions** of the paper are as follows:

1. We devise an advanced approach for the integration of semantic data into programming. This includes on-demand type integration (we rely only on concepts used in the program) based on the theoretical foundations provided by $\lambda_{DL}$, as well as a deep integration of concept expressions and SPARQL queries. This allows for the detection of three kinds of common errors, that occur when working with OWL.
2. We provide a concrete implementation of this approach by extension of the Scala language, including type erasure, integration of an existing reasoner and triple store, while maintaining separate compilability.

However, two important issues are not addressed by the paper. For one, we currently do not provide any form of code completion or general design support dedicated to SPARQL and DL concept expressions. Second, we do not conduct user studies to verify that our approach reduces complexity in practice. Both issues call for future work.

**Road Map** In Section 2, we introduce description logics and SPARQL. In Section 3, we show an essential part of the integration by inferring query types from SPARQL queries. Section 4 describes the essence of integrating DL and SPARQL with typed functional object oriented programming, by extending the syntax and semantics of a formal calculus. In Section 5, we discuss several issues regarding the practical integration of description logics and SPARQL, while Section 6 gives an overview of the architecture and implementation of ScaSpa. In Section 7, we perform a qualitative comparison between our approach and a state-of-the-art library. We focus on how both approaches aid users in dealing with common failure scenarios. This is followed by a discussion of related work in Section 8 and a short summary in Section 9.

## 2 Background

We focus on semantic data formalized in the Web Ontology Language (OWL). Formal theories about OWL are grounded in research on description logics (DL)[1]. Description logics are a family of logical languages used in knowledge representation. They are sub-languages of first-order logic, most of which have been carefully defined to allow for decidable or even PTIME decision procedures.

**Description Logics** A DL knowledge base $\mathcal{K}$ typically comprises two sets of logical axioms: The T-Box (terminological or schematic data) and the A-Box (assertional data). Such axioms are built using the atomic elements defined in the signature of $\mathcal{K}$. The signature provides a set of atomic concept names (e.g., *Person* or *ResearchGroup*), a set of atomic role names (e.g., *worksFor*) and atomic object names (e.g., *bob* or *softlang*).

A role expression is either an atomic relation ($R$) or its inverse ($R^{-}$). Atomic concept names, role expressions and individual objects can be used to built complex concept expressions $C$ using connectives. The available connectives depend on the specific dialect of the description logic. Common connectives include conjunction ($\sqcap$), negation ($\neg$), existential quantification ($\exists$) and enumeration of objects for concept creation. Other connectives (disjunction, universal quantification, ...) may be derived from these. Semantically, a concept is a set of objects. T-Box axioms are constructed from a pair of concept expressions using either the subsumption connective ($\sqsubseteq$) or the equivalence connective ($\equiv$), essentially describing subsumption or equivalences between the various sets of objects. For example, the axiom $\mathit{Employee} \sqsubseteq \mathit{Person} \sqcap \exists \mathit{worksFor}.\mathit{Organization}$ describes that all objects contained in the set *Employee* must also be contained in both the *Person* set, as well as the set of objects for which the role *worksFor* to an *Organization* exists. Assertional axioms on the other hand are either concept assertions or role assertions. A concept assertion $a : C$ claims membership of an object $a$ in a concept expression $C$ (e.g., $\mathit{softlang} : \mathit{ResearchGroup}$ meaning that the *softlang* object is contained in the *ResearchGroup* set). A role assertion $(a, b) : R$ (e.g., $(\mathit{bob}, \mathit{softlang}) : worksFor$) connects objects via role expressions.

Being a subset of first-order predicate logic, DL relies on a Tarski-style *interpretation based semantics*. Axioms contained in either the T-Box or A-Box of $\mathcal{K}$ constitute known facts that must be true. An interpretation in which all facts are true is called a *model*. This may introduce anonymous objects. Consider *alice*, who is a *Chair*. Being a *Chair* requires being the head of a department. However, no department is given for *alice*. This is no inconsistency, but rather incomplete knowledge. An anonymous object is used in the models of $\mathcal{K}$ to represent this department.

An axiom $A$ can be inferred from a knowledge base, written $\mathcal{K} \models A$, if it is true in every model. For example, $\mathcal{K} \models \mathit{ResearchAssistant} \sqsubseteq \mathit{Employee}$ for the $\mathcal{K}$ given in

[1] In practice, OWL is often serialized using RDF. The strict subject-predicate-object triple style of representation introduces some syntactic differences compared to the abstract syntax introduced in this paper.

Figure 1. Importantly, DL relies on an open world assumption. An axiom is true, if it is true in all models. It is false, if it is false in all models. If some models exist in which an axiom is true and some where it is false, then we cannot say whether the axiom is true or false for $\mathcal{K}$. Also, DL does not employ a unique name assumption. Different object names are considered syntactic elements that may semantically refer to the same thing, unless explicitly stated otherwise.

**SPARQL in a DL Context** SPARQL [43] is a graph-matching language built around query patterns. SPARQL supports various entailment regimes, including OWL entailment [20]. While our implementation uses the official SPARQL syntax, in the theoretical parts of the paper we rely on an algebraic formalization for simplicity. We follow [28] in our definitions to focus on constructs that are decidable when used in a DL context. In particular, we restrict ourselves to queries to the A-Box. We also only consider queries returning solution mappings, as opposed to simple boolean `ASK` queries.

A query pattern is an axiom in which at least one object is replaced with a variable. We indicate variables through the meta-variables $x, x_1, x_2$. Therefore, a query pattern *pattern* is defined as follows:

$$pattern ::= x : C \mid (a, x) : R \mid (x, b) : R \mid (x_1, x_2) : R$$

A SPARQL query $q$ is then either a query pattern or the connection of two queries via intersection, union, minus or optional:

$$\begin{aligned} q ::= \;& pattern && \text{(query pattern)} \\ \mid\;& q_1 \;.\; q_2 && \text{(intersection)} \\ \mid\;& q_1 \text{ UNION } q_2 && \text{(union)} \\ \mid\;& q_1 \text{ MINUS } q_2 && \text{(minus)} \\ \mid\;& q_1 \text{ OPTIONAL } q_2 && \text{(optional)} \end{aligned}$$

Formally, a possible solution to a query $q$ is a mapping $\mu$ from (a subset of) variables used in the query onto objects used in $\mathcal{K}$. We write $\mathcal{S}^{\mu}_{\mathcal{K}}[\![q]\!]$ to indicate that $\mu$ is a solution to $q$, given the knowledge base $\mathcal{K}$. This function is defined as follows:

$$\begin{aligned}
\mathcal{S}^{\mu}_{\mathcal{K}}[\![x : C]\!] &= \begin{cases} \mathcal{K} \models a : C & \text{if } \mu(x) = a \\ \text{false} & \text{if } x \notin d(\mu) \end{cases} \\
\mathcal{S}^{\mu}_{\mathcal{K}}[\![(a, x) : R]\!] &= \begin{cases} \mathcal{K} \models (a, b) : R & \text{if } \mu(x) = b \\ \text{false} & \text{if } x \notin d(\mu) \end{cases} \\
&\text{Likewise for } (x, b) : R \text{ and } (x_1, x_2) : R \\
\mathcal{S}^{\mu}_{\mathcal{K}}[\![q_1 \;.\; q_2]\!] &= \mathcal{S}^{\mu}_{\mathcal{K}}[\![q_1]\!] \wedge \mathcal{S}^{\mu}_{\mathcal{K}}[\![q_2]\!] \\
\mathcal{S}^{\mu}_{\mathcal{K}}[\![q_1 \text{ UNION } q_2]\!] &= \mathcal{S}^{\mu}_{\mathcal{K}}[\![q_1]\!] \vee \mathcal{S}^{\mu}_{\mathcal{K}}[\![q_2]\!] \\
\mathcal{S}^{\mu}_{\mathcal{K}}[\![q_1 \text{ MINUS } q_2]\!] &= \mathcal{S}^{\mu}_{\mathcal{K}}[\![q_1]\!] \wedge \neg \mathcal{S}^{\mu}_{\mathcal{K}}[\![q_2]\!] \\
\mathcal{S}^{\mu}_{\mathcal{K}}[\![q_1 \text{ OPTIONAL } q_2]\!] &= \begin{cases} \mathcal{S}^{\mu}_{\mathcal{K}}[\![q_1 \;.\; q_2]\!] & \text{if } (v(q_2) \backslash v(q_1)) \cap d(\mu) \neq \emptyset \\ \mathcal{S}^{\mu}_{\mathcal{K}}[\![q_1]\!] & \text{otherwise} \end{cases}
\end{aligned}$$

where $v(q)$, $d(\mu)$ are the variables occurring in the query, or the mappings domain, respectively. The answer to a query $q$ for a knowledge base $\mathcal{K}$ is $[\![q]\!]_{\mathcal{K}} = \{\mu \mid \mathcal{S}^{\mu}_{\mathcal{K}}[\![q]\!]\}$, the set of all solution mappings $\mu$ for which $\mathcal{K}$ entails the query.

## 3 Type Inference for SPARQL Queries

$$(x : C) : \phi \text{ with } \phi(x) = C \qquad ((x,a) : R) : \phi \text{ with } \phi(x) = \exists R.\{a\}$$

$$((a,x) : R) : \phi \text{ with } \phi(x) = \exists R^{-}.\{a\}$$

$$((x_1,x_2) : R) : \phi \text{ with } \phi(x_1) = \exists R.x_2 \text{ and } \phi(x_2) = \exists R^{-}.x_1 \qquad \frac{q_1 : \phi_1 \qquad q_2 : \phi_2}{q_1 \,.\, q_2 : \phi_1 \odot \phi_2}$$

$$\frac{q_1 : \phi_1 \qquad q_2 : \phi_2}{q_1 \text{ UNION } q_2 : \phi_1 \oplus \phi_2} \qquad \frac{q_1 : \phi_1 \qquad q_2 : \phi_2}{q_1 \text{ OPTIONAL } q_2 : \phi_1 \oplus (\phi_1 \odot \phi_2)} \qquad \frac{q_1 : \phi_1 \qquad q_2 : \phi_2}{q_1 \text{ MINUS } q_2 : \phi_1}$$

$$\begin{aligned}
&\text{where for } (\mathbf{o}, \mathbf{so}) \in \{(\odot, \sqcap), (\oplus, \sqcup)\} \\
&\phi_1 \ \mathbf{o}\ \phi_2 = \{(x, \phi_1(x)\ \mathbf{so}\ \phi_2(x)) \mid x \in dom(\phi_1), x \in dom(\phi_2)\} \\
&\cup \{(x, \phi_1(x)) \mid x \in dom(\phi_1), x \notin dom(\phi_2)\} \\
&\cup \{(x, \phi_2(x)) \mid x \notin dom(\phi_1), x \in dom(\phi_2)\}
\end{aligned}$$

**Figure 2** Rules for concept inference on queries.

In order to provide a typed integration of SPARQL queries into programming, type inference on queries is needed. From a semantics' point of view, a concept expression is a set of values. Queries evaluate to sets of mappings that map variables to values. We therefore infer one concept expression per variable of a SPARQL query. The set defined through the concept expression must at least contain all possible values that a variable can be mapped to after the query has been evaluated. We define the type of a query to be a function $\phi$ mapping each variable in the query to a concept expression.

We use a static analysis of the query through a typing relation $q : \phi$. Query patterns constitute the basic cases of this analysis. In case of a $x : C$ pattern, all possible mappings for $x$ are members of concept $C$. Likewise, for $(a,x) : R$ and $(x,a) : R$, all possible mappings must belong to $\exists R^{-}.\{a\}$ and $\exists R.\{a\}$ respectively. We use $\{a\}$ to denote a so called nominal concept–a concept created by enumerating all its objects. A special case is $(x_1, x_2) : R$. As the concrete concept for $x_1$ is dependent on the concept for $x_2$ and vice versa, we introduce concept references $\exists R.x$ for each of the two variables. These references get resolved after the query has been analyzed. Conjunction, disjunction and *OPTIONAL* in queries are transformed into conjunctions or disjunctions of DL concept expressions in cases where variables are contained in both parts of the query (see the definitions of $\oplus$ and $\odot$ in Figure 2, where *dom* denotes the domain). For *MINUS* queries we have to overestimate the types by disregarding all constraints of the right-hand side: The *MINUS* operator in SPARQL evaluates both operands, before removing all left-hand side solutions incompatible with the right-hand side. Therefore, the overestimation is sound (but a superset of the precise type). This could not be expressed more accurately using concept negation, however, since this notion of negation differs from SPARQL. The complete rules are shown in Figure 2.

As a last step, concept references are resolved. A concept reference $\exists R.x$ is substituted with the respective concept in $\phi$, yielding $\exists R.\phi(x)$. This is repeated until all concept references are eliminated, except possible self references. These cases take the form $\phi(x_1) = \exists R.x_1$ or similar. As we need to replace the reference in a way that captures all possible values, we replace it with the $\top$ concept, yielding $\phi(x_1) = \exists R.\top$. As $\top$ represents the concept containing all objects, this may be a very loose, but sound overestimation.

**Example** Consider the query $((?y, ?x) : \mathit{worksFor} . ?x : \mathit{ResearchGroup})$, selecting tuples of $y$s working for $x$s that are research groups. As a first step, the concept expressions for all patterns are inferred, resulting in the constraint sets $\{y : \exists \mathit{worksFor}.x,\ x : \exists \mathit{worksFor}^{-}.y\}$ and $\{x : \mathit{ResearchGroup}\}$ for the left and right hand side, respectively. Application of the $\odot$ operator yields $\{y : \exists \mathit{worksFor}.x,\ x : \exists \mathit{worksFor}^{-}.y \sqcap \mathit{ResearchGroup}\}$. Resolution of concept references by substitution results in $\{y : \exists \mathit{worksFor}.(\exists \mathit{worksFor}^{-}.y \sqcap \mathit{ResearchGroup}),\ x : \exists \mathit{worksFor}^{-}.(\exists \mathit{worksFor}.x) \sqcap \mathit{ResearchGroup}\}$. Finally, after substitution of self references with $\top$, the following concept expressions are inferred:

x: $\exists \mathit{worksFor}^{-}.(\exists \mathit{worksFor}.\top) \sqcap \mathit{ResearchGroup}$

y: $\exists \mathit{worksFor}.(\exists \mathit{worksFor}^{-}.\top \sqcap \mathit{ResearchGroup})$

## 4 Syntax and Semantics of DOTSpa

In previous work [30], we introduced description logics based types to a simply typed lambda calculus. Here we present syntax and semantics of DOTSpa as extensions to an unspecified formalism. We therefore abstract from specific details, in order to keep DOTSpa as general as possible. In the context of the ScaSpa implementation, however, these definitions can be understood as extensions to the dependent object types calculus (DOT [2]), the theoretical foundation of Scala. In fact, the syntax is a direct extension of the calculus, while the reduction, type assignment and subtyping rules are generalized from the object based nature of DOT. More generally, our approach could easily be transferred to most statically typed, object oriented or functional programming languages, such as F#, C# or Java. Essentially, DOTSpa only requires terms and types, that can be extended with its definitions.

**Syntax** The syntax extension defined by DOTSpa is given in Figure 3. It extends the rules for values, terms and types. Simple values include literals for internationalized resource identifiers (IRI) which, consistent with SPARQL, are used to refer to A-Box instances. Terms are extended by adding the various terms defined by DOTSpa: SPARQL queries and their strictly validated variant, role projections and type case expressions. Strict SPARQL queries use a different validation mechanism than non-strict queries, but are otherwise identical. Role projections query along a single role, providing an easy shorthand notation for this common operation. Type cases are branching expressions, which select one of their branches based on subtyping: They

| | |
|---|---|
| $x, y, z$ | (variable) |
| $i$ | (IRI) |
| $v ::= \ldots$ | (value) |
| $\mid \textbf{iri}\ i$ | (literal IRI) |
| $s, t, u ::= \ldots$ | (term) |
| $\mid \textbf{sparql}\ q$ | (query) |
| $\mid \textbf{strictsparql}\ q$ | (strictly validated query) |
| $\mid t.R$ | (role projection) |
| $\mid t\ \textbf{match}\ \{\ \overline{case}\ \textbf{case}\ _\ \texttt{=>}\ t\ \}$ | (type case) |
| $case ::=$ | (case expression) |
| $\textbf{case}\ x : C\ \texttt{=>}\ t$ | (type case) |
| $S, T, U ::= \ldots$ | (type) |
| $\mid C$ | (concept expression type) |
| $R ::=$ | (role expression) |
| $i$ | (atomic role) |
| $\mid R^{-}$ | (inverse role) |
| $C, D ::=$ | (concept expression) |
| $\{i\}$ | (nominal concept) |
| $\mid i$ | (atomic concept) |
| $\mid \top$ | (top) |
| $\mid \bot$ | (bottom) |
| $\mid \neg C$ | (negation) |
| $\mid C \sqcap D$ | (intersection) |
| $\mid C \sqcup D$ | (union) |
| $\mid \exists R.C$ | (existential quantification) |
| $\mid \forall R.C$ | (universal quantification) |
| $\alpha, \beta ::=$ | (pattern element) |
| $?x$ | (SPARQL variable) |
| $\mid t$ | (term) |
| $pattern ::=$ | (query patter) |
| $\alpha : C$ | (concept assertion) |
| $\mid (i, \alpha) : R$ | (from-lit role) |
| $\mid (\alpha, i) : R$ | (to-lit role) |
| $\mid (\alpha, \beta) : R$ | (role assertion) |
| $q, r ::=$ | (query expression) |
| $pattern$ | (query pattern) |
| $\mid q\ .\ r$ | (intersection) |
| $\mid q\ \textbf{UNION}\ r$ | (union) |
| $\mid q\ \textbf{MINUS}\ r$ | (minus) |
| $\mid q\ \textbf{OPTIONAL}\ r$ | (optional) |

**Figure 3** Syntax extensions defined by DOTSpa.

(RED-ROLE)
$$t.R \rightarrow \textbf{strictsparql}\ (t, ?x) : R$$

(RED-QUERY)
$$\textbf{sparql}\ q \rightarrow \sigma(\llbracket q \rrbracket^*_{\mathcal{K}})$$

(RED-STRICT-QUERY)
$$\textbf{strictsparql}\ q \rightarrow \sigma(\llbracket q \rrbracket^*_{\mathcal{K}})$$

(RED-DEFAULT)
$$v\ \textbf{match}\ \{\textbf{case}\ _\ \texttt{=>}\ t\} \rightarrow t$$

(RED-MATCH)
$$\frac{t \rightarrow t'}{t\ \textbf{match}\ \{\overline{case}\} \rightarrow t'\ \textbf{match}\ \{\overline{case}\}}$$

(RED-CASE-S)
$$\frac{\mathcal{K} \models \{i\} \sqsubseteq C}{\begin{array}{l} i\ \textbf{match}\ \{ \\ \quad \textbf{case}\ x : C\ \texttt{=>}\ t \\ \quad \ldots \\ \quad \textbf{case}\ _\ \texttt{=>}\ u \\ \} \end{array} \rightarrow [x \mapsto i]\ t}$$

(RED-CASE-F)
$$\frac{\mathcal{K} \not\models \{i\} \sqsubseteq C}{\begin{array}{l} i\ \textbf{match}\ \{ \\ \quad \textbf{case}\ x : C\ \texttt{=>}\ t \\ \quad \textbf{case}\ y : D\ \texttt{=>}\ s \\ \quad \ldots \\ \quad \textbf{case}\ _\ \texttt{=>}\ u \\ \} \end{array} \rightarrow \begin{array}{l} i\ \textbf{match}\ \{ \\ \quad \textbf{case}\ y : D\ \texttt{=>}\ s \\ \quad \ldots \\ \quad \textbf{case}\ _\ \texttt{=>}\ u \\ \} \end{array}}$$

**Figure 4** Extended reduction rules.

consist of a term on which cases are matched, the default case as well as zero or more additional cases, restricted with a concept expression.

Types can now be expressed by concept expressions to form concept expression types, using common DL syntax. Additionally, nominal and atomic concepts as well as atomic roles are expressed by IRIs. The remaining rules specify our simplified SPARQL queries (as introduced in Section 2). In this version, however, queries might also contain arbitrary terms in addition to SPARQL variables. This allows the embedding of terms from the language context within a query.

**Semantics** Figure 4 sketches the reduction rules for DOTSpa. In the same manner as for the syntax extension, we specify only rules unique to DOTSpa while omitting rules for simple term reduction.

Role projections (RED-ROLE) are evaluated to equivalent query expressions. An equivalent query for a role projection is the query taking one argument (the term from which the role is selected) and selecting for the particular role. For both strict (RED-STRICT-QUERY) and non-strict (RED-QUERY) queries, the knowledge base (in practice, this would commonly be represented by a SPARQL triple store) has to be consulted to obtain the solution sequence $\llbracket q \rrbracket_{\mathcal{K}}$. For brevity we omit reduction rules for terms embedded in queries. Instead, we assume that terms are reduced via the normal reduction rules by $\llbracket q \rrbracket^*_{\mathcal{K}}$, which is otherwise based on the previously defined $\llbracket q \rrbracket_{\mathcal{K}}$ (Section 2). The query is then mapped to an implementation specific representation via $\sigma$. In terms of evaluation, there is no difference between strict and non-strict queries.

(T-IRI)

$$\Gamma \vdash \textbf{iri}\ i : \{\ i\ \}$$

(<:-CONCEPT)

$$\frac{\mathcal{K} \models C \sqsubseteq D}{\Gamma \vdash C <: D}$$

(T-CASE)

$$\frac{\Gamma \vdash t_n : T_n \qquad \Gamma, x_i : C_i \vdash t_i : T_i \text{ for } i = 1, \ldots, n-1}{\begin{array}{l}\Gamma \vdash i\ \textbf{match}\ \{ \\ \quad \textbf{case}\ x_1 : C_1\ \texttt{=>}\ t_1 \\ \quad \ldots \\ \quad \textbf{case}\ x_{n-1} : C_{n-1}\ \texttt{=>}\ t_{n-1} \\ \quad \textbf{case}\ _\ \ \texttt{=>}\ t_n \\ \} \end{array} \rightarrow lub(T_1, \ldots, T_n)}$$

(T-STRICT-QUERY)

$$\frac{\begin{array}{c} q : \phi \qquad \forall x \in \text{vars}(q) : \mathcal{K} \models \phi(x) \not\equiv \bot \\ \forall t \in \text{terms}(q) : \Gamma \vdash t : C \wedge \mathcal{K} \models C \sqsubseteq \phi(t) \end{array}}{\Gamma \vdash \textbf{strictsparql}\ q : \sigma_T(\phi \text{ where } \forall t \in \text{terms}(q) : \phi(t) \text{ is replaced by } C)}$$

(T-QUERY)

$$\frac{\begin{array}{c} q : \phi \qquad \forall x \in \text{vars}(q) : \mathcal{K} \models \phi(x) \not\equiv \bot \\ \forall t \in \text{terms}(q) : \Gamma \vdash t : C \wedge \mathcal{K} \models \phi(t) \sqcap C \not\equiv \bot \end{array}}{\Gamma \vdash \textbf{sparql}\ q : \sigma_T(\phi)}$$

(T-ROLE)

$$\frac{((t, ?x) : R) : \phi \qquad \forall x \in \text{vars}(q) : \mathcal{K} \models \phi(x) \not\equiv \bot \qquad \Gamma \vdash t : C \wedge \mathcal{K} \models C \sqsubseteq \phi(t)}{\Gamma \vdash t.R : \sigma_T([\phi(?t) \mapsto C]\ \phi)}$$

**Figure 5** Extended type assignment and subtyping rules.

After reducing the matched-on term of type cases (RED-MATCH), the different cases are tried in order: If the value is an IRI and has the respective concept expression type (RED-CASE-S), relying on judgments from the knowledge base, the match evaluates to the respective term with substituted variable. If the matched value does not have the concept expression type (RED-CASE-F), the case is removed. For the single default case, the match expression evaluates to the default expressions term (RED-DEFAULT).

The type assignment and subtyping rules unique to DOTSpa are given in Figure 5. In order to assign the type of match expressions, the least upper bound (lub) of the types of all its branches is used (T-CASE). The lub of concept expression types is defined as the union of concepts ($\text{lub}(C, D) := C \sqcup D$). This definition extends recursively to any arity. IRI values have a nominal concept type, based on the IRI itself (T-IRI). There exists a single subtyping rule for concept expression types: Two concept expression types are in the <: relation, if the respective concepts can be shown to be in a subsumptive relationship in context of the knowledge base (<:-CONCEPT).

In order to type queries (T-QUERY) and (T-STRICT-QUERY), the concept expressions for all variables that occur in a query have to be inferred. Since queries may also contain arbitrary terms, but the algorithm for inference in Section 3 can only deal with SPARQL variables, we map these terms to fresh SPARQL variables before typing the query. Then the mapping $\phi$ can be built according to the inference algorithm. In a slight abuse of notation, we use terms and the fresh variables they map to interchangeably. We also define vars($q$) and terms($q$) to refer to all variables and terms occurring in $q$, respectively. In order to validate a query, all concept expressions inferred for occurring variables $x$ must be satisfiable (i.e., not equivalent to $\bot$). Otherwise, the query can be rejected as always empty. The second validation step varies for the strict and non-strict variants: For strict queries (T-STRICT-QUERY) the concept expression types $C$ of the query-embedded terms $t$ must be subsumed by the inferred types $\phi(t)$ for the matching, freshly introduced SPARQL variables. In the final type, the inferred types are then replaced by the (more specific) concept types $C$. For non-strict queries (T-QUERY) it suffices, that the intersection of $C$ and $\phi(t)$ is satisfiable as well. The final result type is obtained by a function $\sigma_T$, taking the concept expression types as input. The precise type (much like the values constructed by $\sigma$) is not specified for DOTSpa and depends on the implementation. The approach for role projections (T-ROLE) is the same as for strict queries with one argument. This case strongly resembles the behavior of objects and member access. For example, the projection $t.worksFor$ on a term $t$ is evaluated to the query (**strictsparql** $(t, ?x) : worksFor$), meaning it must be proven at compile time that the *worksFor* role exists for the type of $t$.

**Example** Consider again the query (**sparql** $(t, ?x)$ : *worksFor* . $?x$ : *ResearchGroup*) where $t$ is now a term with concept expression type *Employee*. Since this is a non-strict query, validation demands that it is not possible to infer, from the respective knowledge base (Figure 1), that employees may not work for research groups:

$$\exists worksFor.ResearchGroup \sqcap Employee \not\equiv \bot \qquad \text{(T-QUERY)}$$

The non-strict validation approach therefore rejects all queries, where arguments and inferred constraints are known to be disjoint. If, instead of an *Employee*, the argument to this query would have been an *Organization*, the query would have been rejected since only a *Person* can work for something and *Person* and *Organization* are disjoint.

Under strict query validation, it must be possible to prove subsumption between the queries arguments and the inferred constraints:

$$Employee \sqsubseteq worksFor.ResearchGroup \qquad \text{(T-STRICT-QUERY)}$$

The above query is therefore not valid under strict validation, since not all employees work for research groups. If the argument was of type *ResearchAssistant* instead, the query would be valid, since *ResearchAssistant* $\sqsubseteq$ *worksFor.ResearchGroup* is true. In this case, the arguments constraints would be substituted by *ResearchAssistant*, simplifying the type for $?x$ to $\exists worksFor^{-}.ReasearchAssistant \sqcap ResearchGroup$. For common ontologies, this second approach can be too strict a requirement. Therefore, the choice of the validation approach is highly dependent on the respective ontology.

## 5 Instantiating the DOTSpa Framework

DOTSpa is a general language extension framework for introducing querying and a type system for semantic data into programming. We provide a specific implementation called ScaSpa, which implements the DOTSpa approach in the functional programming language Scala. The integration of concept expressions and SPARQL into practical programming technologies, such as the Scala language, introduces several issues. From a practical point of view, the T-Box and A-Box of a knowledge base are often separated. For the T-Box, we rely on ontology reasoners, which are optimized for fast T-Box reasoning. Data however is best stored in a triple store. Both ontology reasoner and triple store are part of ScaSpa in terms of the underlying language integration and architecture.

**Merging of Three Languages** DL concept expressions as well as the SPARQL query language must be syntactically integrated into the host language Scala. We therefore face similar issues as identified by [16, 29, 46], in particular with respect to scoping and the interaction between the languages, such as unquoting of Scala variables in SPARQL queries.

**Knowledge Base Integration into Static Type Checking** DL concept expressions create a new form of types that come with their own set of rules in terms of subtyping, creating an amalgamated type system. The behavior of this new form of types is defined through an ontology reasoner, which must be integrated into the type checking process of Scala, so it can provide judgments to the type checker. This is comparable to the integration of Coq into ML [18].

**Runtime Checks** Objects in a knowledge base do not have a principal type [42] except for the concept that consists only of the object itself. Additionally, knowledge is assumed to be incomplete. Our approach is similar to type-based filters, for example as in LINQ [34] or a type dispatch [19]. However, in our case, such filters or type dispatches require a translation into an equivalent query answered by the triple store.

## 6 Architecture and Implementation of ScaSpa

ScaSpa is a strict extension of Scala. In particular, the type checking process is extended, so that an ontology reasoner can be used for dealing with concept expressions. As this process necessarily relies on typing information, preprocessing in the form of simple desugaring of extended constructs into standard Scala (e.g., by using macros) is not sufficient. Instead, we rely on the extension interface of the Scala compiler. Figure 6 gives an overview of the integration. While we focus on Scala, its primary components can also be understood as a general architecture for transferring DOTSpa into practice.

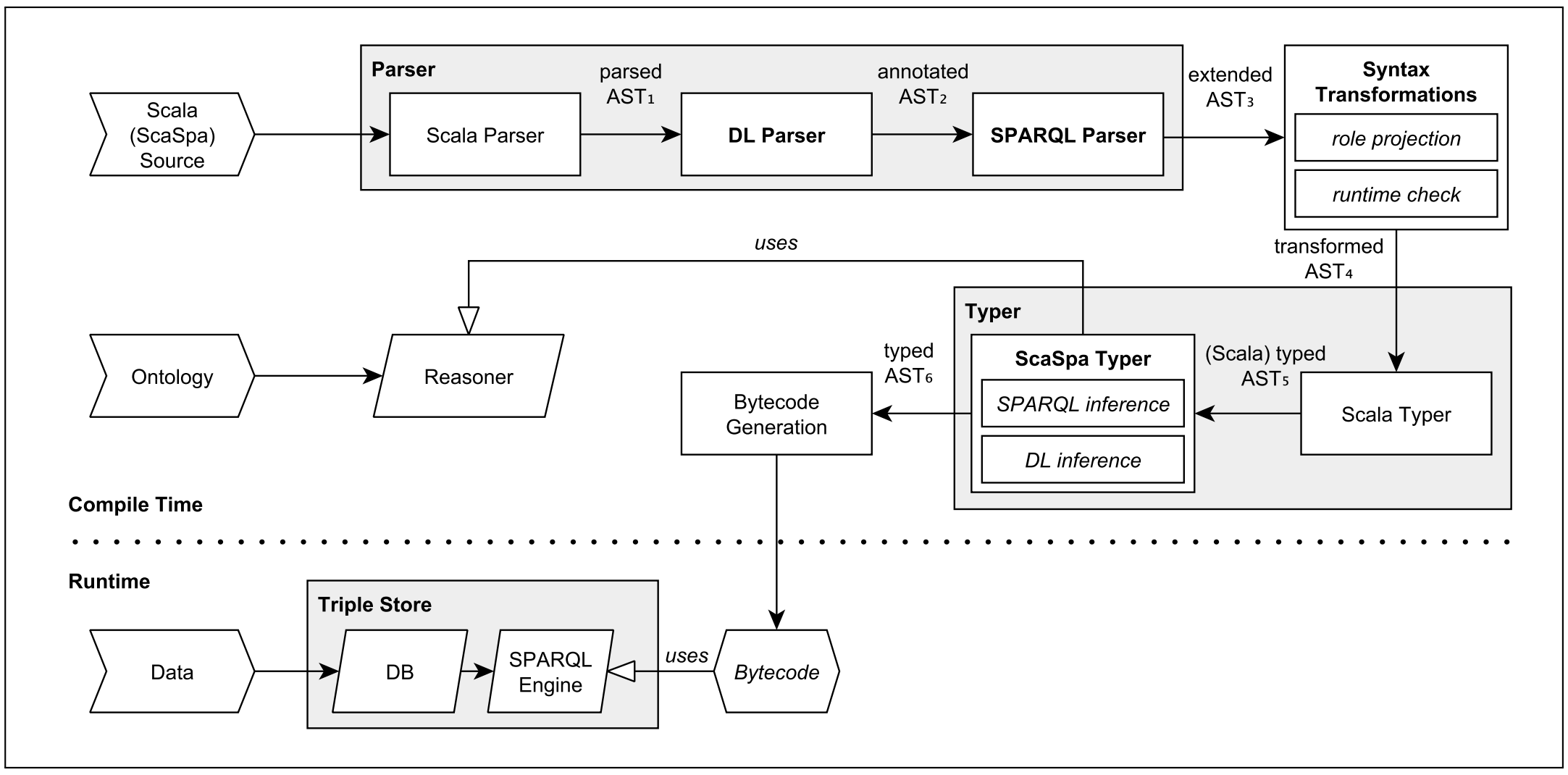

**Figure 6** Architectural model of ScaSpa.
*Nodes are compilation stages (rectangle), summarized stages (shaded rectangle), artifacts (arrow) and external components (parallelogram). Arrows are dataflow (filled heads) and dependency (unfilled heads).*

**Listing 1** Syntax (using back-quotes) and internal representation of concept expression types as static annotations on the base type `DLType`.

```
def empl: `:Person ⊓ ∃:worksFor.:Organization`
// ... transformed to ...
def empl: DLType @dl(":Person ⊓ ∃:worksFor.:Organization")
```

**Parser** We use a staged parsing approach. Initially, DL concept expressions and SPARQL queries are essentially parsed as strings (through the Scala backquote and StringContext features). Syntactic validity of concept expressions and SPARQL queries is ignored at this stage. Instead, the Scala parser creates a standard AST. Later stages recover these constructs through AST traversals: In the DL parser stage, syntactic validity and satisfiability of concept expressions is checked, before erasing them to a base type. The concrete concept expression lives on through a static annotation on this type (see Listing 1). Such static annotations also persist in the (metadata of) the generated byte code, preserving incremental and separate compilability. The concept expressions themselves use standard DL syntax, with the addition that concepts are represented by IRIs. As in SPARQL, prefix aliases can be defined and used. Our examples use the default prefix `':'` for the Lehigh University Benchmark ontology.

Parsing of SPARQL queries, which might contain unquoted Scala expressions, works similarly–the queries are checked for syntactic validity and type annotations are added (see Listing 2). Internally, such a query is represented by the `StringContext` class, which in turn exploits Scala's built-in syntax transformation for prefixed strings. The same feature also handles the insertion of context arguments using `'$'`. For queries,

**Listing 2** SPARQL query syntax relies on the transformation of prefixed Strings (`sparql""`) to instances of `StringContext`.

```
def orgs = sparql"SELECT ?x WHERE { $empl :worksFor ?x }"
// ... transformed to ...
def orgs =
  StringContext("SELECT ?x WHERE {", ":worksFor ?x }")
    .sparql(empl) : List[(DLType, DLType)]
    // Concrete type annotations are inferred at a later stage
```

the parser attaches the general `DLType`, while the specific concept expressions are inferred later, as they might depend on the types of query arguments.

**Syntax Transformations** As a next step, role projections (using member access notation) and type cases are simplified to queries. For role projections, this reduction was already defined in Section 4. Due to the separation of the T-Box and A-Box in reasoner (compile time) and triple store (runtime), runtime subtyping has to be resolved using the triple store. To this end, runtime checks are transformed into an actual `instance-of` test using the base type and a SPARQL `ASK` query–a special form of SPARQL query that evaluates to either true or false. The only limitation of this approach is an over-approximation of results due to the differing notions of negation existing between description logics and SPARQL. This was previously observed for the type inference of *MINUS* queries.

**Typer** After parsing and performing the syntax transformations, the AST contains only valid Scala, including the base type `DLType` with static annotations for concept expression types. This allows the standard Scala typer to do local type inference as well as type checking based on `DLType`. Since there are no more extended constructs, the typed AST is produced in a normal manner. Additional type checker rules for concept expressions and SPARQL queries are implemented in a phase after the Scala typer, relying on the propagation of the base type. The ontology reasoner (ScaSpa uses HermiT [35]) and the actual ontology containing the data descriptions are used during this phase. As OWL includes a namespace feature to distinguish concept expressions, namespace management is also taken care of by the ontology reasoner.

In order to perform type checking and inference according to the rules defined in Section 4, the typed AST is traversed again. During this traversal, the ScaSpa typer performs type checks on, and propagates, the static annotations where the base type was inferred by the Scala typer. A notable difference between the DOTSpa formalization and ScaSpa is that the latter uses a T-Box only mode by default. In T-Box only mode, nominal types are instead estimated using $\top$ (e.g., in literal IRIs) if no explicit annotation is provided. This preserves the separation of T-Box and A-Box. A-Box reasoning can be enabled as well, though the A-Box has to be supplied to the reasoner in this case.

**Listing 3** Least upper bound for concept expression types is the union of concepts.

```
def makeList(prof: `:Professor`, resa: `:ResearchAssistant`) =
  // inferred type : List[:Professor ⊔ :ResearchAssistant]
 List(prof, resa)
```

**Listing 4** Concrete ScaSpa implementation of the $\sigma$ and $\sigma_T$ functions as `List` of n-tuples.

```
def employment: List[(`:Person`, `:Organization`)] =
 sparql"""
  SELECT ?p ?c WHERE {
   ?p :worksFor ?c
  }
 """
```

In addition to the specified typing rules, some further constructs of Scala have to be considered. This includes most importantly type parameters and related features. In order to infer concrete types for type parameters, the least upper bound is sometimes required. This is, for example, the case when inferring the type of if-expressions or methods, for which the same type parameter occurs multiple times (Listing 3). Scala also allows for the explicit definition of variances (invariant, covariant, contravariant) for type parameters, as well as upper and lower bounds. All these features can be directly mapped to the (<:-CONCEPT) rule as defined in Section 4.

Finally, DOTSpa requires implementations to provide a representation for queries, namely the $\sigma$ and $\sigma_T$ functions. In ScaSpa we use simple lists of tuples as demonstrated by the example query in Listing 4.

The AST produced by the ScaSpa typer represents a normal Scala program. Transformation into byte code is therefore a standard procedure. To evaluate queries at runtime, arguments are converted to strings and spliced into the queries. In addition to arguments of concept expression types, ScaSpa supports a limited set of Scala types, which are mapped to appropriate `XSD` data types (e.g., `String` is mapped to `xsd:string`). Therefore it is necessary to take special care and escape these arguments, so that query injections can be avoided. Since queries take only arguments of known types, no syntactical errors can be introduced at runtime. The assembled queries are then handed to a triple store for evaluation. We employ the Stardog [53] triple store in ScaSpa.

**Limitations** Ad hoc polymorphism in the form of method overloading and implicit parameters (used by Scala's notion of type classes), are implemented by compiler internals not exposed through the compiler extension interface. Since all concept expression types are internally represented by the same base type, neither the resolution of overloaded methods nor the implicit search can handle them. Workarounds, such as custom dispatch at runtime or patching the compiler directly, might be possible solutions for this limitation, but remain as future work.

## 7 Comparison with a State-of-the-Art Framework

The primary goal of ScaSpa is to increase type safety and reduce the overall complexity of working with semantic data, by providing an advanced integration in Scala. Here we compare ScaSpa with `banana-rdf` [3], a state-of-the-art Scala library for working with semantic data. The library provides a common interface for its own Scala based back-end (Plantain), as well as support for working with popular Java libraries such as Jena [11]. In order to compare ScaSpa and `banana-rdf`, we identify the following possible failure scenarios that users have to consider when working with SPARQL queries to a concrete knowledge base:

**E-SAT** Query satisfiability. Queries can be unsatisfiable, meaning that there is no possible database instance that can answer the query. If a query is parameterized (e.g., if arguments are spliced into the query), they must be considered as well, to ensure the query remains satisfiable for all possible arguments. Since unsatisfiable queries are always empty, they are meaningless to compute and easily confused with satisfiable queries, that happen to be empty (see also E-EMPTY).

**E-EMPTY** Empty queries. Queries might be empty, even when satisfiable. This occurs, whenever some A-Box data is not explicitly known or simply doesn't exist (e.g., an organization is allowed to have sub-organization, but a particular one might not have any). Empty query results do not constitute an error by themselves, but access to them has to consider the possibility of failure.

**E-SUB** Concept subtyping. Usage of values in a context (T-Box concept), that is not intended by the programmer, e.g., when applying functions to arguments or returning results from functions.

**E-ACCESS** Property access. Access on properties (via role projections) that are not known to exist for a concept.

**E-SYNTAX** Query syntax. Queries can be syntactically invalid, leading to rejection by the triple store and subsequently runtime errors, if not detected at compile time.

In this comparison, we observe how both frameworks aid a user in dealing with these scenarios. As a running example, consider a management application for a university. Such a program may necessitate a function *researchGroups* that, given an *Organization*, lists all *ResearchGroups* that are its direct sub-organizations. This function can be implemented using a simple SPARQL query, splicing in the given organization. Listing 5 shows both a possible implementation of this function using `banana-rdf` (1-11) and ScaSpa (13-20).

**`banana-rdf` approach** The `banana-rdf` implementation of this scenario defines a function *researchGroups* : *Rdf#URI* → `Try`[*Rdf#Solutions*], where the argument is represented by a general type (URI), as is the result (Solutions). Both, when working with the queries result as well as when calling the `researchGroups` function, type safety regarding the concepts in the ontology is not ensured (*not* E-SUB). The SPARQL query for this task is rather simple (lines 5-8) and takes one argument. It has to be constructed as a String value and is parsed at runtime. As a result, syntax errors

**Listing 5** ScaSpa and `banana-rdf` implementations of the `researchGroups` function.

```
// banana-rdf
def researchGroups(org: Rdf#URI): Try[Rdf#Solutions] = for {
  q <- parseSelect(
       s"""
         PREFIX : <http://swat.cse.lehigh.edu/onto/univ-bench.owl#>
         SELECT ?org WHERE {
           ?org a :ResearchGroup .
           ?org :subOrganizationOf <$org>
         }""")
  r <- sparql.executeSelect(q)
 } yield r

// ScaSpa
def researchGroups(org: `:Organization`): List[`:ResearchGroup`] =
 sparql"""
  SELECT ?rq WHERE {
    ?rq a :ResearchGroup .
    ?rq :subOrganizationOf $org.
  }
 """
```

will cause runtime failures and have to be dealt with accordingly (*not* E-SYNTAX). We choose `Try` inside a for comprehension, since it is arguably more elegant than `try-catch` expressions. Neither the spliced-in argument (`org`) nor the query itself are checked for their satisfiability (*not* E-SAT): While the query can be empty for organizations without any (known) sub-organizations (*not* E-EMPTY), this result can therefore not be distinguished from queries which are in fact unsatisfiable to begin with. Worse, since `researchGroups` can be applied to an arbitrary URI, such a failure is not necessarily caused locally (or consistently) by an incorrect query. It might instead only sometimes occur, depending on the specific URI supplied.

**ScaSpa approach** ScaSpa allows the definition of a function *researchGroups* : *Organization* → `List[`*ResearchGroup*`]`, which contains the same, simple SPARQL query with one argument used in the `banana-rdf` solution. Query validation ensures (at compile time) that the query is both syntactically correct (E-SYNTAX) and satisfiable (E-SAT). The types *Organization* and *ResearchGroup* directly represent the concept expressions as defined by the axioms of the ontology (in this case the axioms of Figure 1). This guarantees that the function can only be applied to values which are a subtype of *Organization* as, given the functions specification, intended by the developer (E-SUB). As an added benefit, the type annotations provide a form of documentation, that is guaranteed to remain consistent with the implementation. The type of this argument is considered when checking the queries satisfiability as well. The function is guaranteed to return a list of *ResearchGroups*, which can only be empty, if the A-Box happens to not include any sub-organizations of the given organization, that are research groups (*not* E-EMPTY).

**Listing 6** ScaSpa and `banana-rdf` implementations of the `supervises` function.

```
1  // banana-rdf
2  def supervises(chair: Rdf#URI): Try[Rdf#Solutions] = for {
3    q <- parseSelect(
4          s"""
5            PREFIX : <http://swat.cse.lehigh.edu/onto/univ-bench.owl#>
6            SELECT ?org WHERE {
7              <$chair> :headOf ?org
8            }""");
9    deps <- sparql.executeSelect(q);
10   depit <- Try(deps.iterator.next());
11   dep <- depit("?org");
12   uridep <- dep.as[Rdf#URI];
13   r <- researchGroups(uridep)
14  } yield r
15
16 // ScaSpa
17 def supervises(chair: `:Chair`): List[`:ResearchGroup`] = {
18  val deps = chair.`:headOf`
19  if (deps.nonEmpty) researchGroups(deps.head)
20  else Nil
21 }
```

As a second example, the management application is to be extended by adding a function to determine all research groups subordinate to some *Chair*. In addition, we may want to reuse the `researchGroups` function defined earlier. Listing 6 includes implementations of this function, called `supervises`, using again `banana-rdf` (1-14) and ScaSpa (16-21).

**`banana-rdf` approach** The `banana-rdf` implementation of `supervises` has the same signature as the previously defined function `researchGroups` and is similarly structured. It relies on a SPARQL query, which is again constructed and parsed at runtime (*not* E-SYNTAX), not validated (*not* E-SAT) and then executed. As a next step, the results are accessed: Since query execution returns a general solution mapping, the organization has to be extracted and cast to the appropriate type, involving multiple operations that can fail at runtime (*not* E-SUB). This includes handling the possibility of an empty solution (*not* E-EMPTY). Finally `researchGroups` is called, again without the possibility to perform type checks on either its argument or result (*not* E-SUB), the latter of which is then returned.

**ScaSpa approach** Similar to the first example, ScaSpa allows for the definition of a function with the precise signature $supervises : Chair \rightarrow \texttt{List}[ResearchGroup]$. Since we first need to obtain the organization the given chair is head of, we query for the role `headOf` on the argument `chair`, using role projection. If chairs were not known to be heads of something, this projection would be rejected at compile time (E-ACCESS). Since chairs are heads of departments, the variable `deps` has the inferred concept expression type *Department*. Therefore (and because *Department* is a subtype of *Organization*) we can simply call the previously defined function `researchGroups` to obtain the research groups of the department our chair is the head of (E-SUB). Since the particular department might not be explicitly known (i.e., an anonymous object), we first have to check whether any departments were returned (*not* E-EMPTY). Finally, the return type of `researchGroups` is a subtype of (in fact, equal to) the declared return type of `supervises` (E-SUB).

**Table 1** Summary of the comparison between ScaSpa and `banana-rdf`.

| | ScaSpa | `banana-rdf` |
|---|---|---|
| **E-SAT** | compile time | never |
| **E-EMPTY** | runtime | runtime |
| **E-SUB** | compile time | never |
| **E-ACCESS** | compile time | (*does not apply*[2]) |
| **E-SYNTAX** | compile time | runtime |

**Conclusion** In both scenarios (E-SAT) and (E-SUB), errors are statically avoided at compile time by the ScaSpa compiler. While `banana-rdf` provides some types on the meta-level (such as `Rdf#URI` and `Rdf#Solutions`), it does not aid in the detection of errors in the aforementioned dimensions at all. Access via role projections (E-ACCESS) is also statically checked in ScaSpa, though `banana-rdf` does not include a comparable feature. Syntax errors in SPARQL queries can be detected in both frameworks. However, while ScaSpa rejects invalid SPARQL queries at compile time, queries in `banana-rdf` are constructed and parsed exclusively at runtime, causing runtime failures for malformed queries, which have to be explicitly handled by the user. Finally, a user may have to deal with empty queries (E-EMPTY) in either framework. However, in the case of ScaSpa, users can at least be certain that empty results do not occur due to unsatisfiability of queries. In summary, ScaSpa enables the static detection of three kinds of errors related to the usage of OWL (E-SAT, E-SUB and E-ACCESS), as well as the detection of syntax errors (E-SYNTAX) in queries (Table 1).

[2] Not a feature of `banana-rdf`.

## 8 Related Work

DOTSpa and the ScaSpa implementation are generally related to two larger areas: The language integration of semantic data and technologies, as well as language extension in general.

**RDF and Ontology Integration** The problem of accessing and integrating RDF data in programming languages has been recognized as a challenge in various works. Examples for untyped frameworks include banana-rdf [3], the OWL API [24], Jena [11] and RDF4J [44]. Such frameworks generally provide abstractions on the meta-level, for example in Jena with Java classes such `OntClass` to represent OWL or RDFS classes. While this reflection-like approach might be suitable for developing ontology based tools, it is lacking when working with concrete ontologies [21]. In particular, any correctness of the program related to the data is left completely in the hand of the programmer.

Approaches that create mappings between ontologies and, for example, the object model of object oriented languages, can offer at least some form of verification. Existing mapping frameworks include ActiveRDF [38], Alibaba [52], Owl2Java [27], Jastor [51], RDFReactor [45], OntologyBeanGenerator [1], Àgogo [41] and LITEQ [31]. However, mapping approaches come with their own set of limitations. OWL uses a mixture of nominal and structural typing, and can contain implicit relations (such as the relation between *ResearchAssistant* and *Employee* in Figure 1). For roles in particular, mapping approaches struggle to represent ranges and domains meaningfully. Even in statically typed, mapping-based frameworks, ranges are typically only dynamically checked or use general types and rely on manual casting. The sheer number of potential types can also lead to further problems, including significant runtime overheads. In essence, mappings require the duplication (or approximation) of an ontology reasoner in the target type system.

Some implementations with a deeper integration into programming languages are available. Zhi# [39] extends the type system of C# for OWL and XSD types. The main technical difference is that ScaSpa uses an ontology reasoner in the type checker, allowing for the handling of inferred data. SWOBE [22] provides a typed integration of SPARQL into Java through a precompilation phase–but is limited to primitive data types, IRIs and a triple-based datatype. Additionally, some custom languages exist, that use static type-checking for querying and light scripting to avoid runtime errors [13, 14]. However, the types are again limited in these cases, as they only consider explicitly given statements.

**Language Extension** Extending programming languages is a long standing topic in various domains [16, 29, 46]. Numerous systems, such as TemplateHaskell [47], Racket [17], SugarJ [16], LINQ [34] and Scala macros [10] provide syntactic extensions based on AST transformations. With DOTSpa, we require an extended type checker, so approaches relying solely on transformations of the AST are not suitable. Instead, the conceptual framework proposed in [32] is closer to our approach. Other systems for compiler extensions include Polyglot [36] and ExtendJ [15]. In order to stay within the standard Scala pipeline, rather than creating a new compiler, we rely on Scala compiler extensions instead [48].

ScaSpa relies on an amalgamation of two type systems–one for the normal programming language constructs and one for DL concept expressions. The idea of pluggable type systems [9, 40] that allow for new type systems being layered on top of existing ones has some similarities. Indeed, ScaSpa can be seen as an additional layer on top of the Scala type system. However, the approach is different from ours, since we integrate an ontology reasoner providing the type system judgements. Similarly, open type systems such as provided by the JVM language Gosu [33] allow for the definition of new base types, but do not involve the problem of reasoner integration.

Type providers [12, 31, 50] follow a goal similar to ScaSpa–bridging the gap between programming language and information sources. Compared to more naive mapping approaches, type providers are not directly limited by the size of the data source (types can usually be created lazily, when needed) and do not necessarily cause a runtime overhead, since provided types can be erased. Still, they are afflicted with the other limitations the mapping approach has.

Another related direction in bridging programming language and information source are type systems that are extended for particular kinds of data. Examples include [6] for relational data, [37, 56] for object oriented databases and [5] for XML data. Albeit not being language extensions, the regular expression types provided by CDuce [4] and XDuce [26] are related to ScaSpa due to their unique form of types. Refinement type systems, e.g., provided by F* [49], are somewhat closer to ScaSpa, although typically focused on pre- and postconditions of functions. In contrast, DL concept expressions are logical formulae over nominal and structural type properties. Their defined types are subject to DL reasoning during type checking. As such, ScaSpa is much closer to the integration of Coq in OCaml [18]. In particular to the idea of using the theorem prover, or in our case the ontology reasoner, in the type checking process.

## 9 Summary and Future Work

In this paper we presented DOTSpa, a deep integration of semantic data into practical programming. This is achieved by providing DL concept expressions as a new form of types and via the deep, typed integration of the SPARQL query language. Further, we implement this approach as the ScaSpa extension for Scala. This implementation is based on a staged parsing approach, type judgements provided by an ontology reasoner to the type system, as well as type erasure. We also qualitatively compared our approach with a state-of-the-art Scala library.

Our work can be extended in several directions. Strictly distinguishing between the ontology reasoner at compile time and the triple store at runtime ensures a good runtime performance of match-expressions on concept expressions. As mentioned before, it introduces an overestimation when used in combination with negation. We plan to investigate into performant ways of combining the ontology reasoner and triple store for cases in which negation is involved. Another technical limitation we already mentioned is ad hoc polymorphism. As we erase type information, standard ad hoc polymorphism, such as the method overloading mechanisms provided by Scala, do not work. We plan to investigate possible solutions to this.

As of now, ScaSpa does not provide any support for tooling beyond compilation. Code completion and IDE support are of high interest to us. In particular, code completion on DL concept expressions and SPARQL queries would be useful. Support for tooling opens up another direction of future work: An evaluation using user studies, in order to evaluate the impact of features provided by ScaSpa on real users.

**Acknowledgements** The authors gratefully acknowledge the financial support of project LISeQ (LA 2672/1-1) by the German Research Foundation (DFG).

## About the authors

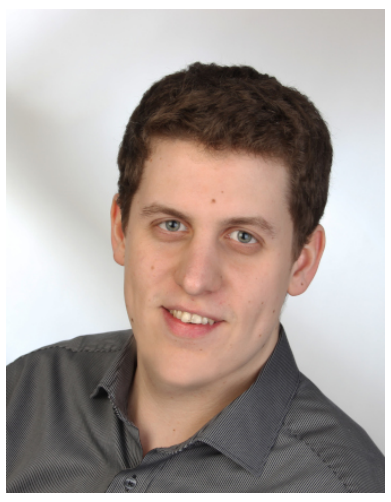

**Philipp Seifer** is a PhD student at the University of Koblenz-Landau. This work is in part based on his master's thesis, with which he graduated from the same university in 2018. He currently conducts his research on the type-safe integration of semantic query languages with programming languages. You can contact him at pseifer@uni-koblenz.de.

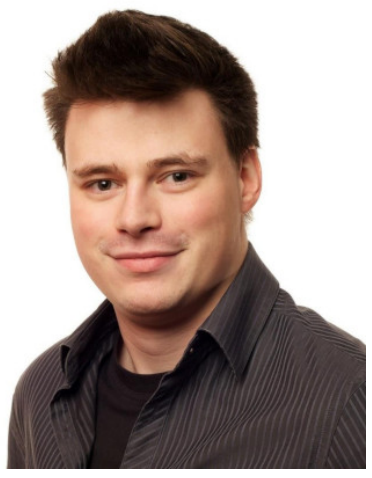

**Martin Leinberger** is a PhD student at the Institute for Web Science and Technologies at the University of Koblenz Landau. He got his masters in 2013 at the same university, and is currently conducting his research over the typed integration of semantic web technologies into programming languages. You can contact him at mleinberger@uni-koblenz.de.

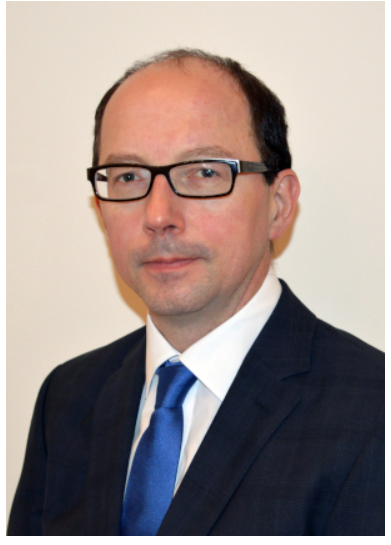

**Ralf Lämmel** is Professor of Computer Science at the University of Koblenz-Landau in Germany and Software Engineer at Facebook in London. His research and teaching interests include software language engineering, software reverse engineering, software re-engineering, mining software repositories, functional programming, grammar-based and model-based techniques, and, more recently, megamodeling. He is author of the textbook "Software Languages: Syntax, Semantics, and Metaprogramming" (Springer 2018). You can contact him at laemmel@uni-koblenz.de.

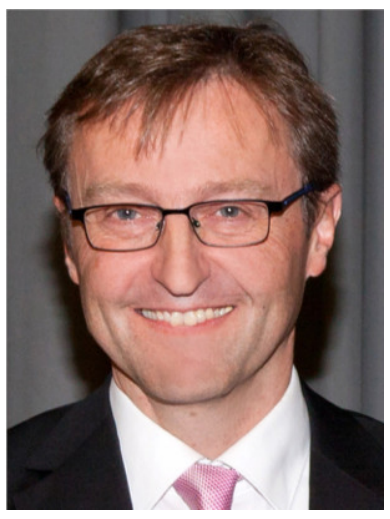

**Steffen Staab** is professor at Universität Koblenz-Landau, Germany, and heads its Institute for Web Science and Technologies (WeST). Steffen also holds a chair for Web and Computer Science at University of Southampton, UK, and he is an associate member of L3S research center at Leibniz Universität Hannover, Germany. He is a fellow of the European Association of Artificial Intelligence and has been awarded the Academy Prize of Rhineland-Palatinate. His interests cover all aspects of semantics, as it concerns data, text, modeling, programming, or user interaction. You can contact him at staab@uni-koblenz.de.